\def\be{\begin{equation}} \def\bea{\begin{eqnarray}}
\def\ee{\end{equation}}\def\eea{\end{eqnarray}}
\def\ncr{\nonumber\\ }
\def\Sch{Schwarzchild }

\def\nn{{g_{00}}} \def\nj{{g_{01}}} \def\jj{{g_{11}}}
\def\nnp{{g_{00}^\prime}} \def\njp{{g_{01}^\prime}} \def\jjp{{g_{11}^\prime}}
\def\rp{{r^\prime}} \def\psip{{\psi ^\prime}} \def\chip{{\chi ^\prime}}
\def\raz{{g_{01}\over g_{11}}}
\def\kr{(\dot r-\raz\rp )}
\def\kpsi{(\dot\psi -\raz\psip )}
\def\kchi{(\dot\chi -\raz\chip )}
\def\kg{(\raz\jjp +\dot\jj -2\njp )}
\def\sg{\sqrt{-g}}
\def\H{{\cal H}} 
\def\MBVR{Maja Buri\'c
\footnote{E-mail: majab@rudjer.ff.bg.ac.yu} and
 Voja Radovanovi\'c\footnote{E-mail: rvoja@rudjer.ff.bg.ac.yu}\\
{\it Faculty of Physics, P.O. Box 368, 11001 Belgrade, Yugoslavia}}

\def\p{\pi}                     

\def\D{\Delta}
\def\F{\Phi}
\def\G{\Gamma}

\def\cl{{\cal L}}

\def\bo{{\raise.05ex\hbox{\large$\Box$}\:}}             
\def\cbo{{\,\raise-.15ex\Sc [\,}}                       
\def\su{\sum}                                           
\def\TH{{\raise.2ex\hbox{$\displaystyle \bigodot$}\mskip-4.7mu \llap H \;}}
\def\face{\hbox{\normalsize$\;\;\:{\raise.9ex\hbox{\oo n}\mskip-13mu \llap
        {${\buildrel{\hbox{\frtnrm ..}}\over\smile}$}}\:$}}     
\def\Face{{\raise.2ex\hbox{$\displaystyle \bigodot$}\mskip-2.2mu \llap {$\ddot
        \smile$}}}                                      
\def\Lhat{{\bf\rlap{\kern-.09em$\hat{\phantom L}$}L}}
\def\Lcheck{{\bf\rlap{\kern-.09em$\check{\phantom L}$}L}}
 
\def\leftrightarrowfill{$\mathsurround=0pt \mathord\leftarrow \mkern-6mu
        \cleaders\hbox{$\mkern-2mu \mathord- \mkern-2mu$}\hfill
        \mkern-6mu \mathord\rightarrow$}
\def\dvec#1{\vbox{\ialign{##\crcr
        \leftrightarrowfill\crcr\noalign{\kern-1pt\nointerlineskip}
        $\hfil\displaystyle{#1}\hfil$\crcr}}}           
\def\ddt#1{{\buildrel {\hbox{\LARGE .\kern-2pt.}} \over {#1}}}
 
\def\frac#1#2{{\textstyle{#1\over\vphantom2\smash{\raise.20ex
        \hbox{$\scriptstyle{#2}$}}}}}                   
\def\sfrac#1#2{{\vphantom1\smash{\lower.5ex\hbox{\small$#1$}}\over
        \vphantom1\smash{\raise.4ex\hbox{\small$#2$}}}} 
\def\bfrac#1#2{{\vphantom1\smash{\lower.5ex\hbox{$#1$}}\over
        \vphantom1\smash{\raise.3ex\hbox{$#2$}}}}       
\def\afrac#1#2{{\vphantom1\smash{\lower.5ex\hbox{$#1$}}\over#2}}    
 
\def\boxes#1{
        \newcount\num
        \num=1
        \newdimen\downsy
        \downsy=-1.64ex
        \mskip-7.8mu
        \bo
        \loop
        \ifnum\num<#1
        \llap{\raise\num\downsy\hbox{$\bo$}}
        \advance\num by1
        \repeat}
\def\boxup#1#2{\newcount\numup
        \numup=#1
        \advance\numup by-1
        \newdimen\upsy
        \upsy=.82ex
        \mskip7.8mu
        \raise\numup\upsy\hbox{$#2$}}
 
\newskip\humongous \humongous=0pt plus 1000pt minus 1000pt

\newif\ifdtup

\def\PRD{Phys. Rev. D\,}
\def\CQG{Class. Quant. Grav.}

\parskip=\medskipamount                 
\def\baselinestretch{1.2}       
\thicklines                         
\def\title#1#2#3#4{
        {\hbox to\hsize{#4 \hfill  #3}}\par
        \begin{center}\vskip.5in minus.1in {\Large\bf #1}\\[.5in minus.2in]{#2}
        \vskip1.4in minus1.2in {\bf ABSTRACT}\\[.1in]\end{center}
        \begin{quotation}\par}
\def\author#1#2{#1\\[.1in]{\it #2}\\[.1in]}

\def\endtitle{\par\end{quotation}\vskip3.5in minus2.3in\newpage}
 
\def\endabstract{\par\end{quotation}
        \renewcommand{\baselinestretch}{1.2}\small\normalsize}
 
\def\start#1{\pagestyle{myheadings}\begin{document}\thispagestyle{myheadings}
        \setcounter{page}{#1}}
 
\catcode`@=11
 
\def\ps@myheadings{\def\@oddhead{\hbox{}\footnotesize\bf\rightmark \hfil
        \thepage}\def\@oddfoot{}\def\@evenhead{\footnotesize\bf
        \thepage\hfil\leftmark\hbox{}}\def\@evenfoot{}
        \def\sectionmark##1{}\def\subsectionmark##1{}
        \topmargin=-.35in\headheight=.17in\headsep=.35in}
\def\ps@acidheadings{\def\@oddhead{\hbox{}\rightmark\hbox{}}
        \def\@oddfoot{\rm\hfil\thepage\hfil}
        \def\@evenhead{\hbox{}\leftmark\hbox{}}\let\@evenfoot\@oddfoot
        \def\sectionmark##1{}\def\subsectionmark##1{}
        \topmargin=-.35in\headheight=.17in\headsep=.35in}
 
\catcode`@=12
 
\def\sect#1{\bigskip\medskip\goodbreak\noindent{\large\bf{#1}}\par\nobreak
        \medskip\markright{#1}}
\def\chsc#1#2{\phantom m\vskip.5in\noindent{\LARGE\bf{#1}}\par\vskip.75in
        \noindent{\large\bf{#2}}\par\medskip\markboth{#1}{#2}}
\def\Chsc#1#2#3#4{\phantom m\vskip.5in\noindent\halign{\LARGE\bf##&
        \LARGE\bf##\hfil\cr{#1}&{#2}\cr\noalign{\vskip8pt}&{#3}\cr}\par\vskip
        .75in\noindent{\large\bf{#4}}\par\medskip\markboth{{#1}{#2}{#3}}{#4}}
\def\chap#1{\phantom m\vskip.5in\noindent{\LARGE\bf{#1}}\par\vskip.75in
        \markboth{#1}{#1}}
\def\refs{\bigskip\medskip\goodbreak\noindent{\large\bf{REFERENCES}}\par
        \nobreak\bigskip\markboth{REFERENCES}{REFERENCES}
        \frenchspacing \parskip=0pt \renewcommand{\baselinestretch}{1}\small}
\def\unrefs{\normalsize \nonfrenchspacing \parskip=medskipamount}
\def\Item{\par\hang\textindent}
\def\Itemitem{\par\indent \hangindent2\parindent \textindent}
\def\makelabel#1{\hfil #1}
\def\topic{\par\noindent \hangafter1 \hangindent20pt}
\def\Topic{\par\noindent \hangafter1 \hangindent60pt}

\documentstyle[12pt,twoside]{article}

\begin{document}

\title{ADM mass of the quantum-corrected \Sch black hole}
{\MBVR}{}{July 1999}

\noindent
We study the hamiltonian and constraints of spherically symmetric
dilaton gravity model. We find the ADM mass of the solution
representing the \Sch black hole in thermal equilibrium with the
Hawking radiation.

\endtitle

\section{Introduction}

Quantum theory of gravity is expected to provide solutions for many
problems of classical general relativity, such as the problem of
singularities, or the problem of interpretation of thermodynamic
quantities like temperature and entropy of the black hole.
As the quantization of gravity is still missing, we do not have the
full quantitative description of these phenomena. Yet, some insights
come from studies of quantization of matter in the curved space and
 semiclassical treatments. 
One of the celebrated achievments on this line was the discovery of
the Hawking radiation \cite{h}, and many papers followed it in
attempt  to describe the backreaction of the radiation
to the  black hole geometry. Also, the Bekenstein-Hawking
formula for entropy \cite{bh} was a subject of many discussions.

The expressions for the temperature of the \Sch black hole $T={1\over
4\p a}$, its entropy $S={\p a^2\over G}$ and energy $E={a\over 2G}$
are all zero'th order or classical expressions and can be derived in
various ways. ($a$  is the radius
of the horizon of \Sch black hole, $a=2MG$.) The problem which is
still open is to find the first quantum corrections to the
above-mentioned quantities, as well as the correction of the metric.
This problem was treated in the literature
in various ways. One approach, done in extensively the eighties \cite{4d}, 
 was to find the expectation value of the
energy-monentum tensor (EMT) of the 
matter field from the symmetry arguments (trace anomaly), and
to solve the coupled Einstein equations for the metric 
\be  R_{\mu\nu}-g_{\mu\nu}{R\over 2}=<\hat T_{\mu\nu}>\quad. \ee
The other possibility is to integrate the matter fields in the path
integral and
determine the effective action $S_{eff}$ to the one-loop order.
 This, unfortunately, has not been done  in four dimensions by now. 
But in two dimensions (2D) this programme has been fulfilled 
for many 2D models.
In the last years there are  various attempts to find
a 2D model which would  successfully describe the 
properties of 4D spherically symmetric solutions and the quantum
corrections of the \Sch solution were examined \cite{fis,no,brm}
Among others, much is expected from the dilaton
spherically symmetric gravity model (SSG).
 In this model, the
quantum  correction is given by  the effective action which is obtained
in \cite{mr4,eff} by evaluation of 2D path integral to the first order
 in $\hbar$.
The 2D classical action is obtained from the 4D Einstein-Hilbert action
which interacts minimally with the scalar field by the spherically
symmetric reduction. The one-loop correction terms are nonlocal,
 but  can be written in the local form after 
introduction of the additional fields $\psi$ and $\chi$ \cite{brm}.  The  fields 
$\psi$ and $\chi$
 are not auxilliary in the usual sense of the Hamiltonian analysis because
they are dynamical, i.e. their equations of motion are of the second
order. This is the remnance of the "quantum origin" of these fields,
i.e. of the fact that they describe the behaviour of the quantized
radiated matter. In this picture,
fixing of the integration constants in the zero'th order
solutions for $\psi$ and $\chi$ corresponds to the choice of the
quantum state of  matter, and it can be done in such a way that the
given solution describes thermalized Hawking radiation (Hartle-Hawking
vacuum).

  In our previous paper \cite{brm}, we obtained the
first quantum correction of the geometry of the \Sch solution,
its temperature  and entropy. We also obtained the value of energy,
assuming that it is defined by the thermodynamic relation $dE=TdS$.
On the other hand, there are known methods for defining energy of the
gravitational field which has a time-like Killing vector and
specified asymptotic behaviour, e.g. the Arnowitt-Deser-Misner
 method \cite{adm}.
As the effective action which we have used in \cite{brm} proved to be
relatively simple in its local form and the quantum corrected
solution is asymptotically flat, it is natural to 
try calculate the ADM mass
of the mentioned solution and compare it to the thermodynamical result.
The problem of finding energy and other conserved quantities in
general relativity is known and well studied \cite{wald}, also in the
context of various 2D theories \cite{bilal,bvv1}. It was applied in
the case of nonlocal potential of the Polyakov-Liouville type in 
the paper of Blagojevi\' c et al \cite{bvv}, and we find their analysis
very instructive for our problem, too.

The plan of the paper is the following: 
 the second section contains the definition of the model, 
the analysis of the hamiltonian
and constraints. The boundary term and
the energy are found in the third section.
 The comparison of the results which are with the thermodynamic ones 
 is given in the
concluding, fourth,  section of the paper.

\section{Hamiltonian and constraints }

As it is well known, the energy-momentum tensor of the gravitational
field is not uniquely defined in general relativity. Also, the value
of energy can be obtained only for some classes of metrics.
If the considered configuration of gravitational field has a time-like
Killing vector, the corresponding conserved quantity can be identified
with the energy of the system if the space is asymptotically Minkowskian.
Also some other classes of metrics allow the identification of
physical time and definition of energy (e.g. asymptotically flat, or
asymptotically de Sitter spaces etc.)
 We will use the Arnowitt-Deser-Misner method 
 in the hamiltonian formulation \cite{wald,bilal,hh}.
 In order to obtain the ADM mass 
  we need to find the hamiltonian and constraints of our system,
  and then, analyzing the variations of hamiltonian,
   find the correct boundary term. Let us first define the
action and the lagrangian we are dealing with.

We start with  Einstein-Hilbert gravity  coupled
minimally to $N$ scalar fields $f_i$ in four dimensions.
This system is described by the action

\be \label{eq:S4}
\G_0=-{1\over 16\p G}\int d^4x\sqrt{-g^{(4)}}R^{(4)}+{1\over
8\pi }\sum _i\int d^4x\sqrt{-g^{(4)}}(\nabla f_i)^2\ ,\ee
After the spherically symmetric reduction
of all fields, we get 2D action
\be  \G _0 = - {1\over 4G}\int d^2 x \sqrt{- g}\left[ e^{-2\F}\left(  R
 + 2 (  \nabla \F)^2 + 2e^{2\F} \right) - 2G\su_{i}
e^{-2\F}(  \nabla f_i )^2 \right]\ .
\ee
Here, $g_{\mu\nu}$ is 2D metric, $\Phi$ is dilaton field, and $f_i$
are the matter fields.
In order to include the quantum effects to the first order in $\hbar$, we add 
to this action the one-loop  quantum correction
which was found in \cite{mr4,eff}.  The effective action which we get is
\bea \label{eq:S2eff} \G&=&-{1\over 4G}\int d^2x\sqrt{-g}
\Bigl( e^{-2\Phi}(R+2(\nabla\Phi)^2+2e^{2\Phi})
-2Ge^{-2\Phi}\sum_i (\nabla f_i)^2 \ncr
&+&
{N\over 8\pi}\int d^2x\sqrt{-g}\Bigl( {1\over 12}R{1\over \Box}R+
R\Phi 
-R{1\over\Box }(\nabla\Phi )^2\Bigr)\ . \eea
 We will
calculate all quantities to the first order in $\kappa$, as the
effective action is also given to this order. 

Since the matter fields enter the action only quadratically and we
are analyzing the correction to the vacuum solution $f_i=0$, we can
introduce $f_i=0$ directly into the action, prior to finding the
equations of motion.
It is convenient to rewrite the effective
action in the local form, using the auxilliary fields
$\psi $ and $\chi$ \cite{brm} and
introducing the field $r=e^{-\Phi}$ instead of $\Phi$.
We get, then
\bea \G&=&-{1\over 4G}\int d^2x\sqrt{-g} \Big[  r^2R+2(\nabla r)^2+2 \ncr
&&-\kappa [2R(\psi -6\chi )+(\nabla\psi )^2-12(\nabla\psi )(\nabla \chi )\ncr
&&-12\psi {(\nabla r)^2\over r^2} - 12 R\log r]\Big] \ . \label{eq:ssga} \eea
Here we introduced the constant  $\kappa ={N\hbar\over 24\pi}$.
The auxilliary fields $\psi$ and $\chi$ satisfy the equations of motion:
\be \Box\psi =R \label{eq:psi} \ee
\be \Box\chi ={(\nabla r )^2\over r^2} \ .\label{eq:chi} \ee
The equations of motion for the other fields are given in \cite{brm}.

The classical part  ($\kappa =0$) of the action (\ref{eq:ssga})
 has the Schwarzschild black hole as
a vacuum  solution.  It reads:
$$f_i=0 \ \ ,\ \ \ r=x^1    \ ,$$
$$g_{\mu\nu}=\pmatrix{-f & 0\cr 0 & {1\over
f}\cr}\ \ ;\ \ \ f=1-{a\over r}\ .$$
$a$ is the radius of the horizon of black hole, $a=2MG$.
The dilaton field $r$ has the role of radius. In the
following, we will  denote $x^0=t$.

The quantum correction of this solution is given 
by the formula  (\ref{solm}-\ref{solc})  and describes the black
hole in  equilibrium with its Hawking radiation.
It was found in   \cite{brm},
 and will be discussed in details in the next section in relation
to the boundary conditions. We will now pass on finding the
hamiltonian corresponding to the action (\ref{eq:ssga}).

One possibility to analyze 2D gravity lagrangians 
 is to fix the gauge partially
and use the lapse and shift functions as  variables \cite{bilal}.
We will proceed along the lines of \cite{bvv}
in order to keep trace of all symmetries. This means that 
for variables we take all components of the metric tensor
 $\nn$, $\nj$, $\jj$, and, along with them,
 $r$, $\psi$ and $\chi$. The conjugated momenta are denoted 
 $\p ^{00}$, $\p ^{01}$, $\p ^{11}$, $\p _r$, $\p _\psi$ and $\p _\chi$.
In order to have only the derivatives of the first order in the lagrangian,
we perform a suitable partial integration. Up to  surface terms
(which are at this stage of the procedure not important and
will be fixed at the end), the lagrangian density corresponding 
to the action (\ref{eq:ssga}) is:
\bea 4G\sg\cl&=&2g +\raz (\dot Q\jjp -Q^\prime \dot\jj )+
\dot\jj\dot Q+\nnp Q^\prime -2\njp\dot Q\ncr
&+&2(1+6\psi {\kappa\over r^2})(\jj {\dot r}^2+\nn\rp ^2-2\nj\dot r\rp )\ncr
&-&\kappa (\jj \dot {\psi }^2+\nn{\psip }^2-2\nj\dot \psi\psip )\ncr
&+&12\kappa (\jj \dot\psi\dot\chi +\nn\psip\chip -\nj
(\dot\psi\chip +\dot\chi\psip )) \ .  \label{eq:ssgl} \eea
Dot and prime denote temporal and spatial derivatives and, to
simplify the expression (\ref{eq:ssgl}),  the function
$Q=r^2+12\kappa\log r-2\kappa (\psi -6\chi )$ is introduced.
The lagrangian 
density (\ref{eq:ssgl})
 does not contain the velocities 
$\dot\nn$ and $\dot\nj$ and therefore
the system is constrained.  Using the definition 
\be \pi _\Phi ={\partial \cl\over \partial\dot\Phi}={\delta L\over\delta\dot\Phi}
 \ ,\ee
 where 
\be L=\int dx^1\,\cl \ ,\ee
 we obtain  the generalized momenta:
\bea  \pi ^{00}&=&0   \label{p00} \\ 
 \pi ^{01}&=&0 \label{p01} \\
 \pi ^{11}&=&  {1\over 4G\sqrt{-g} }\big[ 2rA\kr -2\kappa\kpsi +
12\kappa\kchi \big]  \label{p11} \\
 \pi _r&=&{2\over 4G\sqrt{-g}}\big[ rA\kg +2B\jj\kr \big] \\
  \pi _\chi &=&
 {12\kappa\over 4G\sqrt{-g}}\big[ \kg+\jj\kpsi \big]\ 
 \eea
 \bea
 \pi _\psi ={-2\kappa\over 4G\sqrt {-g}}\big[ 
  ({g_{01}g_{11}^\prime\over g_{11}} +\dot g_{11}-2g_{01}^{\prime})
  +\jj (\dot \psi &-&{g_{01}\over g_{11}}\psi ^\prime  )\ncr
  & -&6\jj\kchi \big] \label{pch}\eea
 $A$ and $B$ are defined as 
$A=1+6{\kappa\over r^2}$, $B=1+6\psi{\kappa\over r^2}$. 

Equations (\ref{p11}-\ref{pch}) can be solved in velocities 
$\dot\jj$, $\dot r$, $\dot\psi$ 
and $\dot\chi$, while the equations (\ref{p00}-\ref{p01})
are primary constraints. The canonical hamiltonian density
$\H = \sum _\Phi \pi _\Phi \dot\Phi - \cl$, obtained from (\ref{eq:ssgl}) is
\bea 
4G\H&=&2\sg +{\sg\over\jj} (2B\rp ^2-\kappa{\psip }^2 +
12\kappa \psip\chip )\ncr
&+&{2\over\sg}\Big({-g\over\jj}\Big)^\prime (Ar\rp -\kappa\psip +6\kappa\chip )\ncr
&+&4G\pi ^{11} (2\njp -\raz\jjp )+4G\raz (\pi _r\rp +\pi _\psi\psip 
+\pi _\chi\chip )\ncr
&+&(4G)^2{\sg\over\jj}\Big( {1\over 144\kappa}\pi _\chi ^2-{\kappa\over 4r^2F}\pi _r^2
-{B\over 2r^2F}(\jj \pi ^{11}-\pi  _\psi )^2\ncr
&+&{1\over 12\kappa}\pi _\psi\pi _\chi +{rA\over 2r^2F}\pi _r
(\jj \pi ^{11}-\pi _\psi  )
\Big)\ ,
 \label{eq:ssgh} \eea
where $F=A^2-{2\kappa\over r^2}B$.

In order to find the secondary constrains, we calculate 
the Poisson brackets of the primary constraints with the hamiltonian, $H=\int
dx^1\,\H$. The Poisson brackets give:
\bea \{  \pi ^{00},H\}&=&-{1\over 2\sg} \H _0\\
 \{ \pi ^{01},H \} &=&{1\over\jj}({\nj\over\sg}\H _0 -\H _1) \ ,\eea
where the secondary constraints $\H _0$ and $\H _1$ are given by
\bea 4G\H _0&=&-{1\over 4G}\Big(2\jj +(2B\rp ^2+\kappa{\psip }^2+12\kappa
\psip\chip )\ncr
&+&4(Ar\rp -\kappa\psip +6\kappa\chip )^\prime -2{\jjp\over\jj}
(Ar\rp -\kappa
\psip +6\kappa\chip )\Big)\ncr
&-&4G\big( {1\over 144\kappa}\pi _\chi ^2-{\kappa\over 4r^2F}\pi _r^2
-{B\over 2r^2F}(\jj \pi ^{11}-\pi _\psi )^2\ncr
&+&{1\over 12\kappa}\pi _\psi\pi _\chi +{rA\over 2r^2F}\pi _r(\jj \pi ^{11}-
\pi _\psi  )
\big)\ , \\
4G \H _1&=&-\pi ^{11}\jjp -2\jj \pi ^{11 \prime}+
\pi _r\rp +\pi _\psi \psip +\pi _\chi \chip \ .\eea
Note that the canonical hamiltonian can be 
written as the sum of the constraints
\be \H =-{\sg\over\jj}\H _0+{\nj\over\jj}\H _1\ ,\ee
which tells us that the only nonvanishing contribution to the energy 
comes from the surface terms which we are to determine. Also note that
the structure of  constraints and hamiltonian is completely analogous
to the one obtained from the Liouville model and PGT \cite{bvv1,bvv}. 
This reflects the fact that all the considered models have the same
symmetries, namely 2D diffeomorphisms. We will not analyse the
symmetry aspects further (algebra of constraints, generators of symmetry),
 but concentrate on on the boundary terms.

Let us review the main idea shortly.
The Hamilton's equations of motion are obtained from the variational principle
\be \delta L=\delta\int dx^1\,(\sum\pi _\Phi\dot\Phi -\H )=0\ ,\ee 
when  the variations are well defined, i.e. when they are  of the form
\be \delta\H =\sum ({\delta\H\over\delta\pi _\Phi}\delta\pi _\Phi +
{\delta\H\over
\delta\Phi}\delta\Phi )\ .\ee
In the case of the hamiltonian densities of the type (\ref{eq:ssgh})
 which contain the spatial derivatives of fields and 
momenta, the terms of the type $\delta\Phi ^\prime$ might occur in
the variation $\delta\H$, and this produces terms
$\delta\Phi\vert_{{\rm bound}}$ in $\delta L$.
In order to make 
the variational procedure consistent,
one adds boundary term to the hamiltonian
to cancel the unwanted variations in 
$\delta L$ and get the Hamilton's equations of motion.. The boundary term 
may not always be defined and its existence depends on the
asymptotic behaviour of the class of the fields 
in which we are performing the variations. This is the point
where the asymptotic behaviour of the fields enters the
definition of the conserved quantities. In the cases where only the
matter fields are varyed the asymptotic conditions are such that the
fields and their derivatives vanish in the asymptotic region and
therefore the boundary term is unimportant. But gravity is not such a
case.

Varying the hamiltonian (\ref{eq:ssgh}), we get
\bea 4G \delta\H &=&{\rm Reg}+\D ^\prime =
{\rm Reg}+\ncr
&+&\Big[ {\sg\over\jj} (4B\rp \delta r-2\kappa\psip\delta\psi 
+12\kappa\psip\delta\chi +12\kappa\chip\delta\psi )\ncr
&+& {2\over\sg}\delta ({-g\over\jj})(Ar\rp -\kappa\psip +6\kappa\chip )+
{2\over\sg}\Big({-g\over\jj}\Big)^\prime (Ar\delta r -\kappa\delta\psi +
6\kappa\delta\chi )\ncr
&+&4G\pi ^{11}(2\delta\nj -\raz\delta\jj )+
4G\raz(\pi _r\delta r+\pi _\psi
\delta\psi +
\pi _\chi \delta\chi )\Big] ^\prime\ , \label{var}\eea
where Reg denotes the regular terms of the type 
$(...)\delta\pi _\Phi +(...)\delta\Phi$. 
We have written explicitely only the terms that give contribution
 on the boundary.
Now we have to examine that contribution for the fields which
asymptotically behave as the \Sch black hole in the Hartle-Hawking vacuum.

\section{Boundary term end energy}

In order to specify the  class of functions i which we are
 varying, let us write the exact solution of the SSG model. 
The static solution is given by:

\bea  r&=&x^1 \label{solr}\\
\nn &=&-f e^{2\Phi}\\
\nj &=&0\\
\jj&=&{1\over f}\\
\psi&=&-\log f - 2\Phi+C\int{dr\over fe^\Phi}
\label{so}\\
\chi&=&{D\over fe^\Phi}+{1\over fe^\Phi}\int{fe^\Phi \over r^2}dr 
 \label{sol} \ .\eea
The dilaton field $r$ plays the same role of
radial coordinate as before.
 The functions $f$ and $\Phi$ are given by
 \bea f(r)&=&1-{a\over r}+\kappa{m(r)\over r}\ ,\\ 
\Phi (r)&=&\kappa (F(r)-F(L))\ ,\eea
where
\bea \label{solm} m(r)&=&{11a\over 4r^2}+{1\over 2r}+{5r\over 2a^2}
+\log{r\over l}\big( {5\over 2a}-{6\over r}+{3a\over r^2}\big)\ ,\\
 F(r)&=& {3\over 4r^2}+{5\over ar}
+\log {r\over l}(-{5\over 2a^2}+{3\over r^2})  \ . \eea
As the functions $\psi$ and $\chi$ enter the hamiltonian always
multiplied by $\kappa$, it suffices to take the zero'th order
solution in their asymptotic behaviour. It reads:
\bea\psi &=&{r\over a}+\log{r\over l}\label{solp}\\
\chi &=&{r\over 2a}-{1\over 2}\log{r\over l} \ .
\label{solc} \eea
(\ref{solm}-\ref{solc}) are written in the form obtained after the
fixing of the integration constants $C$ and $D$ of (\ref{so}-\ref{sol}).
The question of integration constants $C$ and $D$ )
 was analyzed in details in
\cite{brm}. The choice $C={1\over a}$ and $D={1\over 2a}$ which was
taken in (\ref{solm}-\ref{solc}),
 ensures that all functions $\psi$, $\chi$ and
$g _{\mu\nu}$ and therefore also the corrections of the curvature,
energy-momentum tensor, temperature, etc. 
 are regular on the horizon $r=a$, which
is precisely the definition of the Hartle-Hawking vacuum.
The constants $l$ and $L$ have the dimension of length; 
$L$ defines the boundary of the space and is taken to be large, $a\ll L$.

We will perform the variation of the Hamiltonian in the class of the static 
configurations 
with fixed magnitude of space, $L$ and variable mass, $a$.
More precisely, we are considering all configurations
which  asymptotically tend to the given solution i. e.  
differ from it for the terms that decrease like  ${1\over r}$
or faster as $r\to L$. This means that the asymptotic behavior of the 
fields we consider is
\bea  r&=&L+O({1\over L})\ncr
\psi &=&{L\over a}+\log {L\over l}+O({1\over L})\ncr
\chi &=& {L\over 2a}-{1\over 2}\log {L\over l}+O({1\over L})\ .
\eea
The variations and derivatives behave as
\bea \delta r&=&0\ ,\ \ r^\prime =1\ncr
\delta\psi &=&-{L\over a^2}\delta a+O({1\over L})\ ,
\ \ \psi ^\prime = {1\over a}+{1\over L}+O({1\over L^2})\ncr
\delta\chi &=&-{L\over 2a^2}\delta a+O({1\over L})\ ,
\ \ \chi ^\prime = {1\over 2a}-{1\over 2L}+O({1\over L^2})
\ .\eea
 The behaviour of the components of the metric tensor is given by:
\bea \nn &=&-f+O({1\over L})\ncr
\nj &=&O({1\over L})\ncr
\jj&=&{1\over f}+O({1\over L})\ ,\eea
where we have taken
 $e^{\Phi (L)}=1$, as $\Phi (r)=\kappa (F(r)-F(L))$. Also, 
 $\Phi ^\prime = O({1\over L})$, etc. The functions 
$ f(r)=1-{a\over r}+\kappa{m(r)\over r}$ and $m(r)$ for large $r$ 
behave as
\bea m&=&{5\over 2a^2}L+{5\over 2a}\log{L\over l}+O({1\over L})\ncr
\delta m&=&(-{5L\over a^3}-{5\over 2a^2}\log{L\over l}-{1\over
2a^2L})\delta a+
O({1\over L^2})\ncr
f&=&1+\kappa{5\over 2a^2}-{a\over L}+\kappa{5\over 2aL}\log{L\over l}+
O({1\over L^2})\ .
 \eea
 As it can  easily be checked, the given solution has the vanishing momenta,
$\pi _\Phi =0$,  and therefore we can take
\be \pi _\Phi =O({1\over L})\ .\ee

Entering the given behaviour of fields and momenta 
into the  formula (\ref{var})
for the boundary term $\D$  we get
\bea 4G\D&=&(
2f(-\kappa\psip\delta\psi +6\kappa\psip\delta\chi +6\kappa\chip\delta\psi )\ncr
&+&2\delta f(Ar -\kappa\psip +6\kappa\chip )+2f^\prime (-\kappa\delta\psi +
6\kappa\delta\chi ))\Big|_{L}\ .\label{boun}\eea
Leaving only the terms of the highest order in $L$, we obtain
\be 4G\D=-
\delta a \,\Bigl(  2+\kappa ({20L\over a^3}-{12\over a^2}+{5\over
a^2}\log{L\over l})\Bigr)\ . \ee
This, obviously, can be written as a variation of a function 
$H_b$  defined on the boundary: $\D =-\delta H_b$.
This function  is given by
\be  4GH_b=2a+\kappa \Bigl( -{10L\over a^2}+{12\over a}-{5\over
a}\log{L\over l}\Bigr) .\label{Hb}\ee
Note that  the classical limit $\kappa =0$ of  (\ref{Hb}) gives
that $H_b$ is equal to  the mass of the \Sch black hole,
$H_b={2a\over 4G}=M$. Now we can get the complete hamiltonian $H_c$
adding the boundary term $H_b$ to the canonical
hamiltonian $H$. It reads:
\be H_c =\int dx^1 ({\sg\over\jj}\H _0+{\nj\over\jj}\H _1)+H_b\ .\ee
$H_c$ gives the correct equations of
motion because its variation is regular
$$ \delta H_c={\rm Reg}+\int dx^1\D ^\prime -\D={\rm Reg}\ .$$
As we discussed earlier, the fact that $\H _0$ and $\H _1$ are
the constraints  implies that the value of $H_c$ equals to
the value of $H_b$.

Note that the space of the above described quantum corrected
 solution (\ref{solm}-\ref{solc}) is asymptotically flat but
not asymptotically Minkowskian, due to the existence of the Hawking radiation.
The values of the components of metric tensor at infinity are
\bea \nn &=&-f=  -(1+{5\kappa\over 2a^2})+O({1\over L})\\
\jj &=&f^{-1} =1-{5\kappa\over 2a^2}+O({1\over L})\ .\eea
This means that the value of the energy is not simply equal to the
boundary term $H_b$ which we obtained.
In order to find the  
energy  we have to define the coordinates which are 
asymptotically Minkowskian, and and express $H_b$ in that coordinate
system. The new coordinate system is defined by the conditions
$$\tilde g_{00} =-1\ ,\ \ \tilde g_{01} =0\ ,\ \ \tilde g_{11}=1\ .$$
As $g_{00}=({\partial \tilde x^0\over \partial x^0})^2\tilde g_{00}$,
we have ${\partial\tilde  x^0\over \partial
x^0}=\sqrt{-g_{00}}$ and similarly for the 11-component.
The desired transformation in the first
order in $\kappa$ is:
\be \tilde t =(1+{5\kappa\over 4a^2})t \ ,
\ \ \tilde r = (1-{5\kappa\over 4a^2})r\ .\ee
Now we get
the boundary term $\tilde H_b$ in the asymptotically flat coordinates
\be 4G \tilde H_b =2a+\kappa (-{10L\over a^2}+
{19\over 2a}-{5\over a}\log {L\over l})\  \ee
which gives the value of energy. The term proportional to $L$ is the
energy of the hot gas. Note, in the case of the null-dust model, where
the one-loop correction is the Polyakov-Liouville term only, the
boundary term $\Delta$ is given by
\be 4G\Delta=(-2\kappa \psi^\prime \delta \psi +2\delta f(L-\kappa \psi
^\prime)-2\kappa f^\prime \delta \psi )\Big| _{L}\ee
This expression is obtained from (\ref{boun}) by taking $A=B=1, \chi=0.$
It easy to see that the energy is given by    
$$4G\tilde H_b=2a+\kappa ({2L\over a^2}+{1\over 2a}+{1\over a}\log {L\over l})\ .$$
This result is in agreement with \cite{fis,mk}.

\section{Conclusions}

The main conclusion of our calculation is that the value of energy
obtained from the ADM analysis
\be \tilde H_b =M\Bigl( 1-{\kappa \over a^2}({5L\over a}-{19\over
4}+{5\over 2}\log{L\over l})\Bigr) \label{H} \ee
is not equal to the thermodynamic energy which we got in \cite{brm},
\be E=\int TdS=M\Bigl(1-{\kappa\over a^2}({35\over 2}+5\log{L\over
l})\Bigr)\ .\label{E}\ee
Note that the expression (\ref{H}) contains a term
proportional to $L$, while this term is absent from (\ref{E}). 
It is not new that the various definitions of energy in general relativity
do not coincide, even in the classical theory. In the recent paper
\cite{f} Fursaev  show that the hamiltonian of the matter fields
differs from their energy (defined as  $\int T_{00}dV$) for the
spaces which have bifurcate Killing horizons. The matters should be
more complicate when the gravitational energy is also taken into
account. The other reason for the discrepancy might be that the
validity of the thermodynamical approach \cite{fs} might not be
extrapolated beyond the zero'th order level naively. 
Really, it is easy to see that the 
expressions for energy, temperature and entropy obtained in \cite{fis}
for null-dust model and here for SSG 
do not satisfy the thermodynamic relation $dE=TdS$.   
The temperature is obtained by the conical singularity method in both models.
And finally, it
is also possible that the hamiltonian analysis 
does not give the same result when it is applied to 
the local and the nonlocal forms of the same action. In any case,
this question is interesting and deserves further attention and clarification.

\sect{Acknowledgement}
M. B. wants to thank M. Blagojevi\' c for useful discussions
 concerning the ADM mass.

\end{document}